\documentclass{article}
\usepackage{spconf,amsmath,graphicx}

\usepackage{hyperref}

\title{MONAIFBS: MONAI-based fetal brain MRI deep learning segmentation}
\name{Marta B.M. Ranzini, Lucas Fidon, Sebastien Ourselin, Marc Modat, Tom Vercauteren\thanks{The authors are with King's College London, UK.
This work is supported by the Wellcome Trust [WT101957,203148/Z/16/Z], EPSRC [NS/A000027/1,NS/A000049/1]),
the European Union's Horizon 2020 [765148],
and the NIHR BRC at GSTT and KCL. No conflicts of interest.}}
\address{}
\begin{document}
%
\maketitle
%
%
%
In fetal Magnetic Resonance Imaging (MRI), Super Resolution Reconstruction (SRR) algorithms are becoming popular tools to obtain high-resolution 3D volume reconstructions from low-resolution stacks of 2D slices (stacks), acquired at different orientations. 
To be effective, these algorithms often require accurate segmentation of the region of interest, such as the fetal brain in suspected pathological cases.
In the case of Spina Bifida, Ebner \textit{et al.}~\cite{ebner2020} combined their SRR algorithm with a 2-step pipeline composed of a 2D localisation network followed by a 2D segmentation network. However, if the localisation step fails, the segmentation network is not able to recover a correct brain mask. As a result, manual corrections to the brain masks may be needed for an effective SRR.
In this work, we aim at improving the fetal brain segmentation for SRR in Spina Bifida. In particular, we hypothesise that a well-trained single-step UNet can achieve accurate performance, avoiding the need of a 2-step approach. We propose a new tool called MONAIfbs, which takes advantage of the newly developed Medical Open Network for Artificial Intelligence 
(MONAI\footnote{\url{https://monai.io}})
framework for fetal brain segmentation. The proposed model is trained with data from two different centres,
making it more robust to domain shift.

\section{METHOD}
\label{sec:method}

Our network is based on the MONAI dynamic UNet (dynUNet), an adaptation of the nnU-Net framework~\cite{nnUnet2018}.  
Our code and our trained model are made publicly available\footnote{\url{https://github.com/gift-surg/MONAIfbs}}.\\
\textbf{Data\footnote{Anonymized clinical data used in this work were part of a larger research protocol approved by the local ethics committees.}}. Data were available from 2 centres. 604 2D stacks were acquired at Universitair Ziekenhuis Leuven (UZL) (train/validation/test split: 239/42/323); of which, 78 were from healthy cases, the remaining were Spina Bifida. Further 106 2D stacks of Spina Bifida cases were acquired at the Medical University of Vienna (MUW) (77/8/21).\\
\textbf{Network details}. The nnU-Net heuristic rules were adopted to select the optimal network configuration and hyperparameters. 
The dynUNet model was trained  with 2D patches (size [448, 512]) using a combined loss of Dice loss + Cross Entropy loss. Validation was however performed in 3D and the 3D Dice score was used as metric to select the model snapshot best performing on the validation set. Post-processing included selection of the largest connected component. 

\section{RESULTS}
\label{sec:results}
\begin{figure}[t!]
\begin{minipage}[t]{1.0\linewidth}
 \centering
 \centerline{\includegraphics[height=4.3cm]{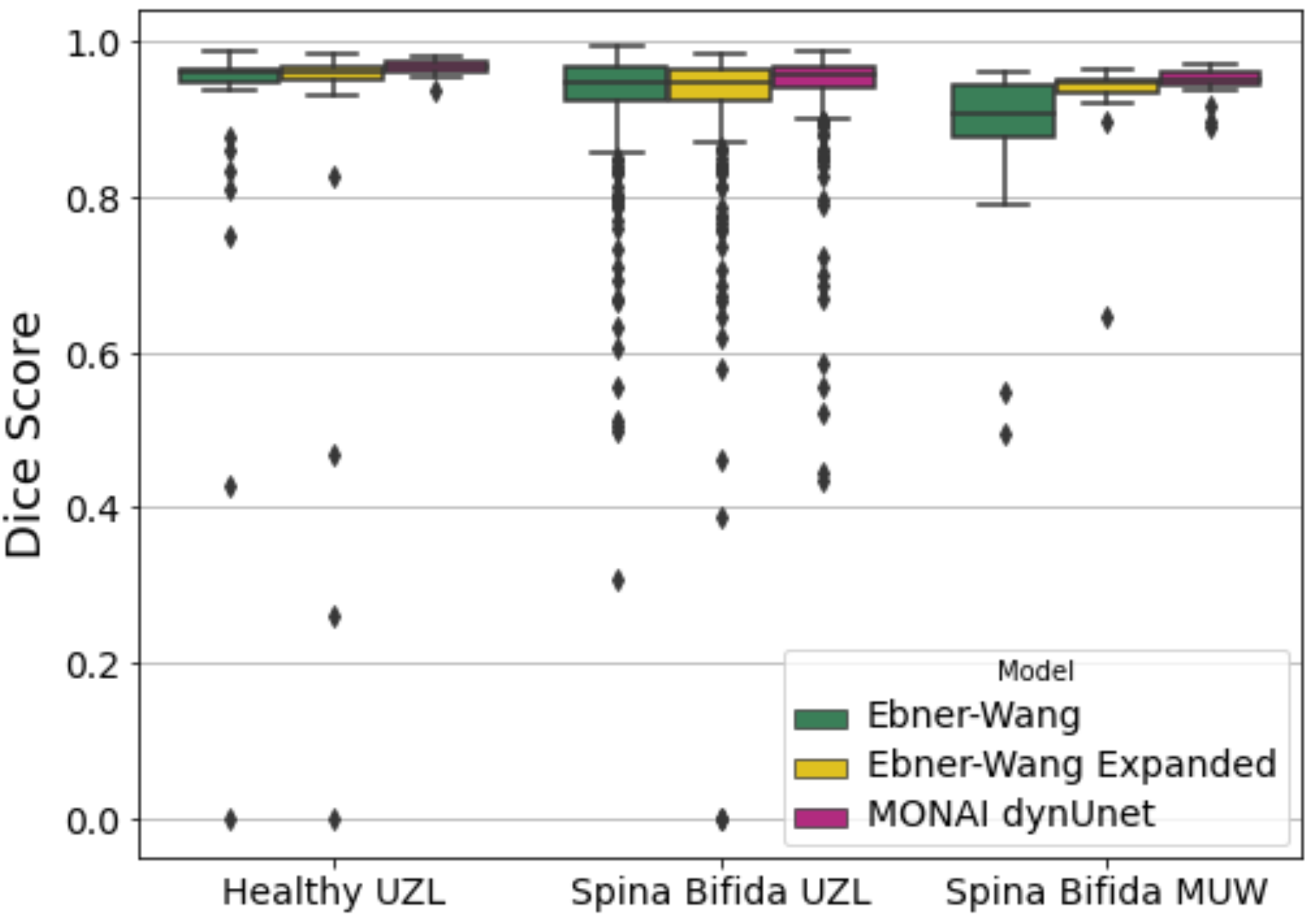}}
\end{minipage}
\caption{Comparison of the proposed MONAI dynUNet model with Ebner-Wang \textit{et al.}~\cite{ebner2020}, original model (in green) and retrained with the expanded dataset (in yellow).}
\label{fig:res}
\end{figure}

Figure \ref{fig:res} shows the dynUNet results on the unseen test set, in comparison to the original two-step approach proposed in Ebner-Wang ~\cite{ebner2020}, and the same Ebner-Wang approach retrained with the expanded dataset available for this work. The dynUNet showed to achieve higher performance using a single step only. It also showed to reduce the number of outliers, as only 28 stacks obtained Dice score less than 0.9, compared to 68 for Ebner-Wang and 53 Ebner-Wang expanded. The proposed dynUNet model thus provides an improvement of the state-of-the-art fetal brain segmentation techniques, reducing the need for manual correction in automated SRR pipelines.

\bibliographystyle{IEEEbib}
\bibliography{bibliography}

\end{document}